\documentclass[onecolumn, 11pt]{revtex4}
\usepackage{amsmath}
\usepackage{graphicx}
\usepackage{subfig}
\usepackage{hyperref}
\usepackage{amssymb}
\usepackage[english]{babel}
\usepackage{epsfig}
\usepackage{wasysym}
\usepackage{graphicx}
\usepackage{indentfirst}
\usepackage{amsmath}
\usepackage{amsfonts}
\usepackage{amssymb}
\usepackage{enumerate}
\begin{document}
%opening
\title{Cosmic string in gravity's rainbow}
\author{ Davood Momeni$^{a}$}
\email{davoodmomeni78@gmail.com}
\author{Sudhaker Upadhyay$^b$}
\email{sudhakerupadhyay@gmail.com}
\author{ Yerlan Myrzakulov$^{a}$}
\email{yerlan83@mail.ru}
 \author{Ratbay Myrzakulov$^{a}$}
 \email{rmyrzakulov@gmail.com}

\affiliation{$^{a}$Eurasian International Centre for Theoretical Physics and\\
Department of General \& 
Theoretical Physics, Eurasian National
University, Astana 010008, Kazakhstan
\\$^{b}$Centre for Theoretical Studies, Indian Institute of Technology Kharagpur, Kharagpur-721302, India}

\begin{abstract}
In this paper, we study the various cylindrical solutions (cosmic strings) in gravity's
rainbow scenario. In particular, we calculate the gravitational field  equations 
corresponding to energy-dependent background. Further, we discuss the possible
Kasner, quasi-Kasner and non-Kasner exact solutions of the field equations.
In this framework, we find that quasi-Kasner solutions can
not be realized in gravity's rainbow. Assuming only time-dependent metric functions, we
also analyse the time-dependent vacuum cosmic strings in gravity's rainbow, which are  completely different than the other GR solutions. 
\end{abstract}
%%%%%%%%%%%%%%%%%%%%%%%%%%%%%%%%%%%%%%%%%%%%%%%%%%%%%%%%%%%%%%%

%\pacs{04.50.Kd, 04.70.Dy.}
%%%%%%%%%%%%%%%%%

%%%%%%%%%%%%%%%%%
\maketitle 
%%%%%%%%%%%%%%%%%%%%%%%%%%%%%%%%%%%%%%%
\section{Overview and motivation}
A strong notion of an observer independent minimum length scale has been found in all theories of quantum gravity, for instance,  in string theory \cite{amati}, noncommutative geometry \cite{girelli}, loop quantum gravity 
\cite{rovelli, carlip} and Lorentzian dynamical triangulations 
\cite{card, ambjorn1, ambjorn2, ambjorn3}. Here we point out that the nascent GW astronomy \cite{abo} could help in discriminating among  general relativity or alternative theories   \cite{cor}.  There is no harm to assume this minimum measurable length scale as the Planck scale. The mathematical ground of general  theory of relativity is based on a smooth manifold which  breaks down when   energies of probe
reaches the order of Planck energy \cite{Maggiore,Park}. Keeping this point in mind, one 
may expect a radically new
picture of spacetime, which includes departure from the standard
relativistic dispersion relation. A departure from the standard dispersion relation
indicates that the system incorporates a breaking of Lorentz invariance.  
Indeed, Lorentz symmetry is one of the most remarkable symmetries in nature which along with the Poincar\'e symmetry fix   the standard form of energy ($E)$-momentum ($\vec{p}$) dispersion relation, i.e., $E^2-|\vec{p}|^2=m^2$. 
A modification in the standard 
energy-momentum dispersion relation occurs in the ultraviolet limit of most of the  quantum gravity theories \cite{Hooft, Kostelecky, Amelino-Camelia1,
Gambini, Carroll}. In fact,   a modification in the energy-momentum dispersion relation is studied in Horava-Lifshitz gravity   in the ultraviolet region \cite{horava1, horava2}. 
Although the broken Lorentz invariance is considered  in ultraviolet limit,   the velocity of light $c$ and the Planck energy $E_p$ should not be modified. The study of such modified energy-momentum dispersion relation (MDR)  is known as
double special relativity (DSR) (or
  non-linear special relativity) \cite{Magueijo-smolin1,Magueijo-smolin2}.

An extension of DSR into a general
relativity framework which has at its foundation the proposal that the geometry of
a spacetime  runs  with the energy scale at which the geometry is being probed  is known as gravity's rainbow  (or might be called  ``doubly general relativity") \cite{Magueijo-smolin3}.
In this regard, they found that the cosmological distances, in an expanding
universe, become energy dependent. In fact, by considering the energy dependent time, 
they addressed  the
horizon problem  without inflation or a varying speed of light  \cite{Magueijo-smolin3}. 
Earlier, it has been found that the gravity's rainbow produces a deformation to the spacetime
metric which becomes significant at  the Planck scale of  the particle's energy/momentum. 
Moreover, it is realized that  quantum corrections can become
relevant not only for particles approaching the Planck energy but, due to the one loop contribution,
even for low-energy particles as far as Planckian length scales are considered \cite{rem}.
The   gravity's
rainbow  illustrates  a new mass-temperature relation and define
a minimum mass and maximum temperature for rainbow
black hole predicting the existence of black hole remnant \cite{ling0,ling00}. It has also been found that
the gravity's rainbow prevents
black holes from evaporating completely \cite{ling,ahmed}, just like the standard
uncertainty principle prevents the hydrogen atom from collapsing \cite{ad,ad1}.
The quantum corrections due to rainbow functions of the metric are studied in thermodynamics of the massive BTZ black holes \cite{sud}.
It has been found  there that in semi-classical/quantum regime, thermodynamics of the black holes would be
modified into a level which differs from classical case. More precisely, the different orders of the rainbow functions
affect the high energy and asymptotical behaviors of the solutions and their leading terms  \cite{sud}.

Moreover, the gravity's rainbow has also been studied at various occasions  in recent past.   For instance, the critical behavior of  the black holes in Gauss-Bonnet gravity's
rainbow was discussed and   found that the generalization to a charged case puts an energy dependent restriction on different parameters \cite{hend}.
By considering rainbow functions  in terms of
power-law  of the Hubble parameter, the Starobinsky
model of inflation, from the  perspectives of  gravity's rainbow,  was investigated  \cite{cha}.   
Very recently, the modifications on Hawking-Page phase transition \cite{z,y}  and wave function of the universe \cite{m}   are also discussed. 
In gravity's rainbow framework, the Hawking, Unruh, free-fall and fiducial  
temperatures of the black hole  have also been investigated \cite{yad,gim}.
 In addition,  remnants of black objects \cite{fai},  asymptotic flatness \cite{jon},  nonsingular universes in Einstein and Gauss-Bonnet gravities
\cite{adel,adel1} have been studied in the gravity's rainbow background.  
The zero point energy in a spherically symmetric background combining the high energy distortion of gravity's rainbow with the modification induced by a $f(R)$ theory
has been interpreted in Ref. \cite{rem1,pana1}. Within the context of gravity's rainbow modified geometry, motivated from quantum gravity corrections at
the Planck energy scale,  it is shown that the distortion of the metric leads to a Wheeler-DeWitt equation
whose solution admits outgoing plane waves and consequently, a period of cosmological inflation may arise without the need of introducing an inflation field \cite{rem2}.
%%%%%%%%%%%%%%%%%%%%55
In cosmology of early Universe, we investigate the generally accepted doctrine that the universe is affecting to what we termed as "topological defects"  through exhaustion of all sources of matter, and suggest that by virtue of a cosmic string mechanism which maintains its available energy is self-gravitating. Energy is being "degraded" in objects which are in the cosmos, but "elevated" or raised to a higher level in strings \cite{topological_defects1,topological_defects2}.
 The modified Friedmann-Robertson-Walker(FRW) equations are also derived in
 the contexts of gravity's rainbow  \cite{ling1}. 
From the astrophysical perspective, it was shown that the existence of energy dependent spacetime can modify the hydrostatic equilibrium equation (or modified TOV) of stars \cite{pana}.
 One main motivation for us to study exact cylindrical solutions in gravitational theories is to describe such topological defects by Riemannian geometry. A simple description of the above topological defects is to find the cylindrical solution by solving highly non linear field equations.  In general relativity (GR), the simplest cylindrical model described by the  class of exact cylindrical solutions were found by Kasner and later on studied by several authors \cite{Kasner1}
-\cite{Tian}.

On the other hand, the spontaneous symmetry breaking mechanism and cosmological phase transitions together urged us to confront the possibility of topological defects 
 playing a significant role in cosmology \cite{ve}.  The cosmic strings, in fact,
 provide a viable fluctuation spectrum for galaxy formation \cite{ze}.
 The study of topological defects and cosmological implications of strings are subjects of
 sustained interest \cite{hind}.
The  cylindrical  solutions play a major role  in the study 
gravitational systems. For instance,
  cylindrical  solutions describe the gravitational waves    by an effective energy tensor, in terms
of a gravitational potential generalizing the Newtonian potential \cite{sea}.
Also,   cylindrical symmetry gets relevance in order to  study the exact solutions
in general relativity  (not only due to  the theoretical reasons but also for the physical realization in objects such as cosmic strings).  The cylindrical  solutions   are
discussed from the  viewpoint of exact solutions in $f (R)$ gravity theories \cite{aza}.
Although the substantial progress has been made in this area but  the cosmic string
solutions   in gravity's rainbow framework are still unexplored. This provides  
an opportunity to us to bridge this gap.  
%%%%%%%%%%%%55
%%%%%%%%%%%%%%%%%%%%%%%%%%5
One of the oldest branch in 
 GR   is to find  exact solutions of certain types of gravitational theories as Remannian metrics $g_{\mu\nu}$ satisfy some types of  field equations.  Recently, exact solution finds different interesting applications in other complex problems of physics \cite{exact-sol}. A celebrated application is found when  exact forms for a type of gravity can be used to probe the quantum theory on the associated spatial boundary. Generally speaking to proceed with exact solutions, we need to have two basic choices: first option is to fixing symmetry of the background metric  $g_{\mu\nu}$; second is to  give the symmetry to the matter contents $T_{\mu\nu}$. Although, in gravitational theory, there 
 is no simple and direct relation between the symmetry of source of the gravitational field and the symmetry of the metric because of non-linearity of field equations and breaking of linear approximations, but still we can probe symmetry very carefully from matter. As we know, topology is an independent parameter and can be imposed on geometry after we fix the general form of metric. It  also becomes  important when some types of the  topological effects are needed  suddenly by assigning an independent metric.
%%%%%%%%%%%%555
In the context of modified theories of gravity, cosmic strings are investigated
 in $f(R)$ gravity \cite{aza, f(R)_strings*}, teleparallel theories \cite{Torsion_strings, Teleparallel_strings, Houndjo:2012sz}, brane worlds \cite{Braneworld_strings}, Kaluza-Klein
models \cite{KK_strings}, Lovelock Lagrangians\cite{Lovelock_strings}, Gauss-Bonnet
\cite{GB_strings,Rodrigues:2012qu,Houndjo:2013us}, Born-Infeld \cite{BI_strings1,
BI_strings2}, bimetric theories\cite%
{Bimetric_strings}, non-relativistic models of gravity \cite{Misc_strings},   scalar-tensor theories \cite{ScalarTensor_strings1,ScalarTensor_strings2,ScalarTensor_strings3,ScalarTensor_strings4,
ScalarTensor_strings6,ScalarTensor_strings7},   Brans-Dicke theory \cite{Delice:2006uz,Baykal:2009vs,Delice:2006gs,Baykal:2005pv,Kirezli:2012vw,Ciftci:2015cua}, dilation gravity \cite{Dilaton_strings1,Dilaton_strings2}, non-minimally coupled models of gravity \cite{Harko:2014axa}, Mimetic gravity \cite{Momeni:2015aea}
and, recently, in the  Bose-Einstein condensate strings \cite{Harko:2014pba}. However, the
 cosmic string is still unexplored in gravity's rainbow setting. Here, we try to 
 bridge this gap.

%%%%%%%%%%%%%5
In this paper, we first briefly review the basics of energy-dependent Einstein field equations  described by a  specific rainbow functions. Following the basic properties of
cosmic strings, we write the metric of the cosmic string in gravity's rainbow.
Implementation the metric of the cosmic string in gravity's rainbow to  the Einstein 
field equations leads to a set of differential equations in terms of rainbow function.
In order to realize the   exact solutions of these differential equations, we consider the  two parametric metric solution (so-called Kasner  solution, 
 which is an unique exact solution for the Einstein equations with cylindrical symmetry).
In addition to that we also discuss the possibility of  the quasi-Kasner and non-Kasner solutions. In this regard, we find that the quasi-Kasner solutions can
not be realized in gravity's rainbow. In this setting, we further compute the Ricci   and Kretschmann scalars  and observe that the gravity's rainbow cosmic strings have same   singularities as in the standard GR 
theory. We discuss the time-dependent cosmic strings also in gravity's rainbow.
The time-dependent vacuum solutions are  based on  the assumption that all metric functions depend on time only not on space.

We organize this work as follows. In section II, we discuss the basic set-up
gravity's rainbow. The  equations of motion for cosmic string in gravity's rainbow
is computed in section III. The realization of Kasner's solution in   gravity's rainbow
is given in section IV. We try to emphasize the spherically symmetric solution for 
gravity's rainbow by considering energy as a function of radial coordinate only in section V.
The time dependent cosmic string solutions are discussed in 
section VI. The behavior of cosmic string in gravity's rainbow is discussed by considering
energy as a function of time  only in section VII. We summarize  results in the last section.  
 
%%%%%%%%%%%%%%%%%%%%%%%%%%%%%%%%5
\section{Basic set-up of gravity's rainbow } 
 Gravity's   rainbow  (doubly general relativity) is an extension of DSR into a general relativity framework, which justifies a modified dispersion relation  given by \cite{Magueijo-smolin1,Magueijo-smolin2}
\begin{eqnarray}
E^2f^2\left(E/E_p\right) - |\vec{p}|^2 g^2\left(E/E_p\right) = m^2c^4,
\end{eqnarray}
where $E_p$ refers the Planck energy. Here, functions $f\left(E/E_p\right)$ and  $g\left(E/E_p\right)$ are known as the  rainbow functions. This modification in the energy-momentum relation due to rainbow functions becomes significant  in the ultraviolet limit. However, the following constrained are required  to reproduce the standard dispersion relation in the infrared limit:
\begin{eqnarray}
 \lim\limits_{E/E_p\rightarrow 0}f\left(E/E_p\right) = 1 ~; \lim\limits_{E/E_p\rightarrow 0}g\left(E/E_p\right) = 1 ~.
\label{rafu}
\end{eqnarray}
In order to express the energy-dependent metrics in a one-parameter family, we write \cite{Magueijo-smolin3}
\begin{eqnarray}
g(E) = \eta^{ab}e_a\left(E/E_p\right) \otimes e_b\left(E/E_p\right)
\label{rafu1}
\end{eqnarray}
where  the energy-dependent set of orthonormal frame fields $e_a = (e_0, e_i)$ are
\begin{eqnarray}
e_0\left(E/E_p\right) = \frac{1}{f\left(E/E_p\right)}\tilde{e}_0,\ \ \ \  e_i\left(E/E_p\right) = \frac{1}{g\left(E/E_p\right)}\tilde{e}_i~.
\label{rafu2}
\end{eqnarray}
Here, the quantities  ($\tilde{e}_0, \tilde{e}_i$) are  the energy independent frame fields.
Here, we also note that, in the limit $E/E_p \rightarrow 0$, this corresponds to usual general relativity.
 Eventually, these  gravity's rainbow functions 
modify the black hole metrics. Motivated from loop quantum gravity considerations \cite{alfaro,sahlmann,smolin},  our analysis is based on the following specific rainbow functions \cite{Amelino-Camelia1}:
\begin{eqnarray}
f(E/E_p) = 1~ ;~ g(E/E_p) = \sqrt{1-\eta \left(\frac{E}{E_p}\right)^n}
\label{rainbow functions}
\end{eqnarray}
where $\eta$ refers to the rainbow parameter. We follow the natural units $c=
\hbar=k_B=1$ throughout the paper.  
 
%%%%%%%%%%%%%%%%%%%%%%%%%%%%%%%%%%%%%%%%%%%%%%%%%%%%%%%%%%
%%%%%%%%%%%%%%%%%%%%%%%%%%%%%%%%%%%%%%%%%%%%%%%%%%%%%%%%%% 
 
\section{Metric and equations of motion}
In this section, we compute the equations of motion for cosmic string in gravity's rainbow.  
This type of cylindrical solution has treated popular in the literature for some time with the name of cosmic string as a model to describe topological defects of early cosmology and closed time like curves.
%%%%%%%%%%%%%%%%%%%%%%%%%%%%%%%%%%%%

According to standard definition, the general cosmic string  metric has the following basic properties:
\begin{eqnarray}\label{rho'}
&&g_{\mu\nu}(t,r,\varphi,z)=
\left\{
\begin{array}{lr}
g_{ta}=0\ , & a=\{r,\varphi,z\} \\
\partial_{t}g_{\mu\nu}=0
\ , &\ \ \ \ \ \ \mu,\nu=\{t,r,\varphi,z\}
\\
\partial_z,\partial\varphi
\ , & \mbox{symmetries}
\\
\mathcal{R}^3\times \mathcal{S}^1
 , & \mbox{topology}%
\end{array}%
\right. \ .
\end{eqnarray} 
Inevitably, the cosmic string with cylindrical symmetry describes an exact solution in 
general relativity with  a metric which presents an interior solution when   
radius tends to zero.
%%%%%%%%%%%%%%%%%%%%%%%%%%%%%%%%%5

The cosmic string metric $g_{\mu\nu},\ \ \mu=\{t,x^a\},\ \ a=1,2,3$,  is supposed to be  static (i.e with vanishing off diagonal  components $g_{ta}=0,a=\{r,\varphi,z\}$ and time independent $\partial_{t}g_{\mu\nu}=0$) and  cylindrical symmetric. It means we suppose that the distribution of matter fields in such space time forms  a static stress-energy tensor. By symmetry, we will think about existence of a pair of commuting killing vectors  $\partial_z,\partial\varphi$. From kinematical point of view, the  classical trajectories (orbits) are closed  around the $z$ axis. The unique coordinate to be used to describe  the metric $g_{\mu\nu}$ is the (semi) radial coordinate $r$ (because it doesn't mean distance in general). This coordinate $r$ is initiating  from the $z$ axis when $r=0$ and it is supposed to be extended smoothly up to the spatial infinity $r=\infty$. By definition, the topology of the space time is uniquely well defined by the $\mathcal{R}^3\times \mathcal{S}^1$, here $\mathcal{R}$ denotes  the domain of real numbers and $\mathcal{S}^1$ defines a unit circle.

%%%%%%%%%%%%%%%55
Let us start by writing the metric of the time-independent cosmic string in gravity's rainbow scenario as follows,
\begin{equation}
ds^{2} = \frac{A(r)}{f^{2} \left(E/E_{p}\right)} dt^{2} - \frac{1}{g^{2} \left(E/E_{p}\right)}dr^{2} - \frac{B(r)}{g^{2} \left(E/E_{p}\right)} d \varphi^{2} - \frac{C(r)}{g^{2} \left(E/E_{p}\right)} dz^{2},\label{metric1}
\end{equation}
where $A(r), B(r), C(r)$ are some functions depend on cylindrical coordinate $r$.

The Einstein field equation obtained by varying the usual Einstein-Hilbert action with respect to the rainbow's metric $g_{\mu\nu}\left(E/E_{p}\right)$ is given as follows:
\begin{eqnarray}
R \left(E/E_{p}\right)_{\mu\nu}-\frac{1}{2}g \left(E/E_{p}\right)_{\mu\nu}R&=&8\pi G \left(E/E_{p}\right)T_{\mu\nu}.\label{feq}
\end{eqnarray}
whereas the energy dependent Newton's universal gravitational constant 
$G\left(E/E_p\right)$ becomes the conventional Newton's universal gravitational constant $G = G(0)$ in the limit  $E/E_p\rightarrow0$.
We assume that the matter fields fill the spacetime with energy-momentum tensor $T_{\mu}^{\nu}=\mbox{diag}(\rho,-p_r,-p_{\varphi},-p_z)
$, where $\rho$ is energy density and $p_i$ corresponds to different components of pressure field.
%%%%%%%%%%%%%%%%%%5
By substituting (\ref{metric1}) in field equation (\ref{feq}), we obtain the following set of ordinary differential equations:
\begin{eqnarray}
\frac{B^{'2}}{B^2}-\frac{B^{'}C^{'}}{BC}+\frac{C^{'2}}{C^2}-\frac{B^{''}}{B}-\frac{C^{''}}{C}&=&32\pi G \left(E/E_{p}\right)\frac{\rho}{g^2},\label{A2}\\
\frac{A^{'}B^{'}}{AB}+\frac{A^{'}C^{'}}{AC}+\frac{B^{'}C^{'}}{BC}&=&32\pi G \left(E/E_{p}\right)\frac{p_r}{g^2},\label{B2}\\
 \frac{A^{'2}}{A^2}-\frac{A^{'}C^{'}}{AC}+\frac{C^{'2}}{C^2}-2\frac{A^{''}}{A}+2\frac{C^{''}}{C}&=&-32\pi G \left(E/E_{p}\right)\frac{p_\varphi}{g^2},\label{C2}\\
 \frac{A^{'2}}{A^2}-\frac{A^{'}B^{'}}{AB}+\frac{B^{'2}}{B^2}-2\frac{A^{''}}{A}-2\frac{B^{''}}{B}&=&-32\pi G \left(E/E_{p}\right)\frac{p_z}{g^2}\label{D2}.
\end{eqnarray}
In order to discuss the behavior of these equations, we need to  them.
We emphasize this in forthcoming sections.

\section{ Realization of Kasner's solution in gravity's rainbow }\label{kasner}
Here, to find the exact solution, we focus, in particular,   Kasner's solution  in gravity's rainbow. 
The  two parametric metric, so-called Kasner solution, is an unique exact solution for 
 the  Einstein equations with cylindrical symmetry in GR
\cite{Kasner1,Kasner2,exact-sol}. This is given by following line element:
\begin{eqnarray}  \label{Kasner1}
ds^{2} = (kr)^{2a}dt^{2} - dr^{2} - \beta^{2} (kr)^{2(b-1)}r^{2}d{\varphi}^2-(kr)^{2c}dz^{2}, \label{Kasner2} 
\end{eqnarray}
here  $k$ defines an appropriate length scale and $\beta$ is a constant  and it is related directly  to the
deficit angle of the conical space-time.
By solving  Einstein  equations (\ref{feq}), we are considering  not only the form of metric functions for which $R_{\mu\nu}=0$  but also every possible value of the parameters $\{a,b,c\}$ satisfying,
\begin{equation}
a + b + c=a^2 + b^2 + c^2=1. 
\end{equation}
In Kasner metric, Ricci scalar vanishes (i.e. $R=0$), however,   for quasi-Kasner  solution one can 
have non-vanishing Ricci scalar (i.e. $R\neq0$). We will show that  the Kasner metric is  a trivial solution in  gravity's rainbow.
A possible question will be, ``does the quasi-Kasner solution  with $R\neq 0$ solves our gravity's rainbow system described by the equations   Eqs. (\ref{A2}-\ref{D2}) or not"?. We address this problem in the following situations:
%%%%%%%%%%%%%%%%%%
\par

\begin{itemize}
\item  \it{Quasi-Kasner solutions in Rainbow scenario}: $A=(kr)^{2a},\ B=\beta^{2}r^{2} (kr)^{2(b-1)},\ C=(kr)^{2c}$:
\end{itemize}

The condition $T_{\mu\nu}\neq0$  in (\ref{A2}-\ref{D2}) would not harm us  to find  the quasi-Kasner's solutions. It will be interesting enough to find something similar to GR solutions. Substituting these values of solutions  in Eqs. (\ref{A2}-\ref{D2}), we observe that for particular values of parameters $\big(a,b,c\big)$  the quasi-Kasner is a solution 
for field equations in gravity's rainbow. These are 
\begin{eqnarray}
&&ds^{2} = \frac{dt^{2}}{f^{2} \left(E/E_{p}\right)} - \frac{dr^{2}}{g^{2} \left(E/E_{p}\right) }- \frac{\beta^{2} k^{2}d{\varphi}^2}{g^{2} \left(E/E_{p}\right)}-
\frac{(kr)^{2}dz^{2}}{g^{2} \left(E/E_{p}\right)}.\\&&
ds^{2} = \frac{dt^{2}}{f^{2} \left(E/E_{p}\right)} - \frac{dr^{2}}{g^{2} \left(E/E_{p}\right)} - \frac{\beta^{2}r^{2}d{\varphi}^2}{g^{2} \left(E/E_{p}\right)}-
\frac{dz^{2}}{g^{2} \left(E/E_{p}\right)}.\\&&
ds^{2} = \frac{(kr)^{2}dt^{2}}{f^{2} \left(E/E_{p}\right)} - \frac{dr^{2}}{g^{2} \left(E/E_{p}\right)} - \frac{\beta^{2} k^{2}d{\varphi}^2}{g^{2} \left(E/E_{p}\right)}-
\frac{dz^{2}}{g^{2} \left(E/E_{p}\right)}.
\end{eqnarray}
From the above expressions, it is obvious that if we put the values of parameters $\big(a,b,c\big)$  into the metric (\ref{Kasner1}), we obtain a class of Kasner metrics   with  $R=0$. This implies that  the quasi-Kasner solutions can not be realized in  gravity's rainbow.
\begin{itemize}
\item \it{ Non-Kasner family of the exact solutions}:
\end{itemize}
In order to study the non-Kesner type of solution,  we first eliminate  $A$, $C$ and $B$ respectively from  Eqs. (\ref{A2}-\ref{D2}) and, as a result, we obtain:
\begin{eqnarray}
&&{\it B'''}=-{\frac {{{\it B'}}^{2}{\it B''}-2\,{{\it B''}}^{2}B}{{\it
B'}\,B}}\label{B}
\\&&
{\it C''}=-{\frac {2\,{\it B''}\,B{\it B'}C-{{\it B'}}^{3}C+{\it B''}
\,{B}^{2}{\it C'}}{{\it B'}\,{B}^{2}}}\label{C}
\\&&{\it A'}=-{\frac {{\it B'}\,{\it C'}\,AB  -2\,{{\it B'
}}^{2}AC+4\,{\it B''}B\,AC}{- B{\it C'}
 ^{2}+ B' B C}}\label{A},
\end{eqnarray}
along with the following constraint:
\begin{eqnarray}
-B'^2BC'C-B'^3C^2+B'B^2C'^2+2B''B^2BC'+2B''B'BC^2 = 0.
\end{eqnarray}
Now, it is possible to  solve these three differential equations given in (\ref{B}-\ref{A}). In this way, we found the following exact solutions for system of equations:
\begin{eqnarray}
 &&A(r)=l_3\exp\left[{-l_1\int  dr \frac{ {l_4}({l_1}+\frac{n}{3}-2)
   (r-r_0){}^{\frac{n}{2}}+3 {l_5}({l_1}-\frac{n}{3}-2) ( r-r_0 ){}^{-\frac{n}{2}}}{
{l_4}
   ({l_1}-\frac{n}{3}-\frac{2}{3})
   ( r-r_0 ){}^{1+\frac{n}{2}}+{l_5}
   ({lC_1}+\frac{n}{3}-\frac{2}{3})
   ( r-r_0 ){}^{1-\frac{n}{2}}} }\right],
\label{Asol}
\\
&&B(r)=\tilde{l}_3(r-r_0)^{l_1},
\label{Bsol}
\\&&C(r)={ l_4}( r-r_0)^{-\frac{1}{2}{l_1}+1+\frac{1}{2}n}+{l_5}\,
( r-r_0) ^{-\frac{1}{2}{l_1}+1-\frac{1}{2}n},\label{Csol}
\end{eqnarray}
where $\tilde{l}_3=l_3 e^{i\pi l_1},n=\sqrt{-3l_1^2+4l_1+4},l_3>0$ , $l_1\in[-\frac{2}{3},2]$,$n\in \mathcal{R}$ and $r\geq l_2$. 
 
 Here we note that for $n=0$  the metric converts to  the following  form:
\begin{eqnarray}
 ds^2=\frac{\rho^2 d\tilde{t}^2}{f^{2} \left(E/E_{p}\right)}- \frac{d\rho^2}{g^{2} \left(E/E_{p}\right)}-\frac{\mu  d\varphi^2}{g^{2} \left(E/E_{p}\right)}-  \frac{d\tilde{z}^2}{g^{2} \left(E/E_{p}\right)}, 
 \label{sol2} 
\end{eqnarray}
where we defined $\mu=l_4+l_5,\ \rho=r-l_2,\ \tilde{t}= t\sqrt{\tilde{l}_3},\   \tilde{z}=z\sqrt{l_3}$.
In fact, this metric  corresponds to the vacuum Levi-Civita  metric   and  it coincides with the cosmic string.
Here, we also mention that  the azimuthal angle $\varphi\neq \in(0,2\pi]$, but the geometry  still remains  close to flat space. The  deficit angle,  in this case, is    $1-4\eta\left(E/E_{p}\right)=\mu\left(E/E_{p}\right)$, where $\eta\left(E/E_{p}\right)$ is the gravitational mass per unit length of the spacetime.\par
We compute Ricci scalar and get
\begin{eqnarray}
&&R= g^{2} \left(E/E_{p}\right)\Big[e^{i\pi(1+\,{l_1})} n^2{
l_1}\,{ l_4}\,{ l_5}\Big]\rho ^{-1+2\,{l_1}}.
\end{eqnarray}
Here we notice that the Ricci scalar $R\neq0$.  The  Kretschmann scalar $\mathcal{K}=R_{\mu\nu\alpha\beta}R^{\mu\nu\alpha\beta}$  corresponding to this metric 
% is given by
%\begin{eqnarray}
%&&\mathcal{K}=\frac{D(\rho,l_1,..,l_5)\Big({l_4} {\rho}^{n/2}+{ l_5}\,{
%\rho}^{-n/2} \Big)^{-1}}{{\rho}^{4} \Big({l_4}\, \left( -3\,{ l_1}+n+2
% \right) {\rho}^{n/2} -  {l_5}\left( 3\,{l_1}+n-2 \right)\,{\rho
%}^{-n/2} \Big) ^{4}  }.
%\end{eqnarray}
%\end{widetext}
%Here we find that $D(\rho,l_1,..,l_5)$ 
is a non-singular function for $\rho\in\mathcal{R}$, however, the  Kretschmann scalar has singularity at $\rho=0,\ \rho=\left(\pm i\sqrt{\frac{l_4(-2+3l_1 -n)}{l_5(3l_1+n-2)}}\right)^{-2/n} (\mbox{with}\ l_5 \neq 0),\ \rho=\exp \left(-\frac{1}{n}\log \frac{l_4}{l_5} \right) (\mbox{with}\ \frac{l_4}{l_5} > 0)$.
Thus, we can  conclude that gravity's rainbow cosmic strings  have same  singularities as in the standard GR theory. 
%%%%%%%%%%%%%%%%%%%%%%%%%%%%%%%%%%%%%%%%%%
\section{When $E=E(r)$}
In this section, we consider the general $f(R)$ case,
where  modified Einstein equations become \cite{aza},
\begin{eqnarray}
FR_{\mu\nu}-\nabla_\mu\nabla_\nu F=\frac{1}{4}g_{\mu\nu}(FR-\square F),
\end{eqnarray}
where $F(R)=\frac{df(R)}{dR}$.
From above, we can write 
\begin{eqnarray}
\frac{FR_{\mu\mu}-\nabla_\mu\nabla_\mu F}{g_{\mu\mu}}=\frac{1}{4}(FR-\square F 
=\frac{1}{4}g_{\mu\nu}(FR-\square F):=A_\mu. \label{aa}
\end{eqnarray}
This means that the combination  $A_\mu$
  is independent of the
index $\mu$ and therefore $A_\mu= A_\nu$ for all $\mu, \nu$.
Now, in  case $E=E(r)$, the metric (\ref{metric1}) becomes
\begin{equation}
ds^{2} = \frac{A(r)}{f^{2} \left(E(r)/E_{p}\right)} dt^{2} - \frac{1}{g^{2} \left(E(r)/E_{p}\right)}dr^{2} - \frac{B(r)}{g^{2} \left(E(r)/E_{p}\right)} d \varphi^{2} - \frac{C(r)}{g^{2} \left(E(r)/E_{p}\right)} dz^{2}.\label{metric11}
\end{equation}
The nonzero components of the metric tensor have the following expressions:
$$
g_{00} = \frac{A(r)}{f^{2} \left(E(r)/E_{p}\right)} , \  g_{11} = - \frac{1}{g^{2} \left(E(r)/E_{p}\right)}, \  g_{22} = - \frac{B(r)}{g^{2} \left(E(r)/E_{p}\right)}, \  g_{33} = - \frac{C(r)}{g^{2} \left(E(r)/E_{p}\right)}.
$$
It is easy to find the inverse of the above components as
$$
g^{00} = \frac{f^{2} \left(E(r)/E_{p}\right)}{A(r)}, \  g^{11}= - g^{2} \left(E(r)/E_{p}\right), \ g^{22} = - \frac{g^{2} \left(E(r)/E_{p}\right)}{B(r)}, \  g^{33} = - {\frac{g^{2} \left(E(r)/E_{p}\right)}{C(r)}}.
$$
With these  metric components, it is straightforward to calculate the Christoffel symbols $\Gamma_{\nu\delta}^{\mu}$ with following definition: 
\begin{equation}
\Gamma_{\nu\delta}^{\mu} = \frac{1}{2} g^{\mu\theta} \left(\frac{\partial g_{\theta\nu}}{\partial x^{\delta}} + \frac{\partial g_{\theta\delta}}{\partial x^{\nu}} - \frac{\partial g_{\nu\delta}}{\partial x^{\theta}}\right).\label{chr}
\end{equation}
The  calculation leads to the  expressions of nonzero components as
\begin{eqnarray}
\Gamma^{0}_{10}&=&\frac{A'}{2 A}-\frac{f_EE'}{fE_p},\ \Gamma^{1}_{00}=\frac{g^2 A'}{2 f^2}-\frac{A g^2 f_E E'}{f^3 E_p}, \ \Gamma^{1}_{11}=-\frac{g_E E'}{g E_p}, \ \Gamma^{1}_{22}=-\frac{1}{2} B^{'}+\frac{B g_E E'}{hE_p},\\ \nonumber
\Gamma^{1}_{33}&=&-\frac{1}{2} C^{'}+\frac{C g_E E'}{g E_p},\ \Gamma^{2}_{21}=\frac{1}{2} \frac{B^{'}}{B}-\frac{g_E E'}{g E_p},\ \Gamma^{3}_{31}=\frac{1}{2} \frac{C^{'}}{C}-\frac{g_E E'}{g E_p}.
\end{eqnarray}
The Ricci tensor is defined by
\begin{eqnarray}
R_{\mu\nu}&=&\frac{\partial\Gamma^\delta_{\mu\nu}}{\partial x^\delta}-\frac{\partial\Gamma^\delta_{\mu \delta}}{\partial x^\nu}+\Gamma^\delta _{\mu\nu}\Gamma^\theta_{\delta\theta}-\Gamma^\theta_{\mu \delta}\Gamma^\delta_{\nu \theta}.\label{ri}
\end{eqnarray}
With the help of above definition, the different components are calculated by
\begin{eqnarray}
R_{00}&=&\frac{1}{2}\frac{g^2}{f^2}A^{''}+\frac{g^2 A' B'}{4 B f^2}+\frac{g^2 A' C'}{4 C f^2}-\frac{g^2 A^{'2}}{4 A f^2}-\frac{g A' g_E E'}{2 f^2 E_p}-\frac{g^2 A' f_E E'}{f^3E_p}-\frac{A g^2 B' f_E E'}{2 B f^3 E_p}-\\&& \nonumber
\frac{A g^2 C' f_E E'}{2 C f^3E_p}+\frac{2 A g^2 f_E^2 E^{'2}}{f^4 E_p^2}-\frac{A g^2 f_{EE} E^{'2}}{f^3E_p^2}-\frac{A g^2 f_E E^{''}}{f^3 E_p}+\frac{A g f_E g_EE^{'2}}{f^3 E_p^2}, \\
R_{11}&=&-\frac{A''}{2 A}+\frac{A' f_E E'}{A fE_p}-\frac{A' g_E E'}{2 A g E_p}+\frac{A^{'2}}{4 A^2}-\frac{B^{''}}{2 B}+\frac{B' g_E E'}{2 B hE_p}+\frac{B^{'2}}{4 B^2}-\frac{C^{''}}{2 C}+\frac{C' g_E E'}{2 C g E_p}+\\&& \nonumber
\frac{C^{'2}}{4 C^2}+\frac{f_{EE} E^{'2}}{fE_p^2}+\frac{f_E g_E E^{'2}}{f g E_p^2}+\frac{f_E E^{''}}{f E_p}-\frac{2 f_E^2 E^{'2}}{f^2 E_p^2}+\frac{2 g_{EE} E^{'2}}{g E_p^2}+\frac{2 g_E E^{''}}{g E_p}-\frac{2 g_E^{2} E^{'2}}{g^2 E_p^2},\\
R_{22}&=&-\frac{A' B'}{4 A}+\frac{B A' g_E E'}{2 A g E_p}-\frac{B^{''}}{2}-\frac{B' C'}{4 C}+\frac{B' f_E E'}{2 f E_p}+\frac{B' g_E E'}{g E_p}+\frac{B^{'2}}{4 B}+\frac{B C' g_E E'}{2 C g E_p}-\\&& \nonumber
\frac{B f_E g_E E^{'2}}{f g E_p^2}+\frac{B g_{EE} E^{'2}}{g E_p^2}+\frac{B g_E E^{''}}{g E_p}-\frac{2 B g_E^{2} E^{'2}}{g^2 E_p^2},\\
R_{33}&=&-\frac{A' C'}{4 A}+\frac{C A' g_E E'}{2 A g E_p}-\frac{B' C'}{4 B}+\frac{C B' g_E E'}{2 B g E_p}-\frac{C^{''}}{2}+\frac{C' f_E E'}{2 f E_p}+\frac{C' g_E E'}{hE_p}+\frac{C^{'2}}{4 C}-\\&& \nonumber
\frac{C f_E g_E E^{'2}}{f g E_p^2}+\frac{C g_{EE} E^{'2}}{g E_p^2}+\frac{C g_E E^{''}}{g E_p}-\frac{2 C g_E^{2} E^{'2}}{g^2 E_p^2}.
\end{eqnarray}
The mixed Ricci tensor is computed as
\begin{eqnarray}
R_{0}^{0}&=&\frac{g^2 A^{''}}{2 A}+\frac{g^2 A' B'}{4 A B}+\frac{g^2 A' C'}{4 A C}-\frac{g^2 A' f_E E'}{A f E_p}-\frac{g A' g_E E'}{2 A E_p}-\frac{g^2 A^{'2}}{4 A^2}-\frac{g^2 B' f_E E'}{2 B f E_p}-\\&& \nonumber
\frac{g^2 C' f_E E'}{2 C f E_p}-\frac{g^2 f_{EE} E^{'2}}{f E_p^2}-\frac{g^2 f_E E^{''}}{f E_p}+\frac{g f_E g_E E^{'2}}{f E_p^2}+\frac{2 g^2 f_E^2 E^{'2}}{f^2 E_p^2}, \\
R_{1}^{1}&=&\frac{g^2 A^{''}}{2 A}-\frac{g^2 A' f_E E'}{A f E_p}+\frac{g A' g_E E'}{2 A E_p}-\frac{g^2 A^{'2}}{4 A^2}+\frac{g^2 B^{''}}{2 B}-\frac{g B' g_E E'}{2 B E_p}-\frac{g^2 B^{'2}}{4 B^2}+\\&& \nonumber
\frac{g^2 C^{''}}{2 C}-\frac{g C' g_E E'}{2 C E_p}-\frac{g^2 C^{'2}}{4 C^2}-\frac{g^2 f_{EE} E^{'2}}{f E_p^2}-\frac{g^2 f_E E^{''}}{f E_p}-\frac{g f_E g_E E^{'2}}{f E_p^2}+\\&& \nonumber
\frac{2 g^2 f_E^2 E^{'2}}{f^2 E_p^2}-\frac{2 g g_{EE} E^{'2}}{E_p^2}-\frac{2 g g_E E^{''}}{E_p}+\frac{2 g_E^2 E^{'2}}{E_p^2},\\
R_{2}^{2}&=&\frac{g^2 A' B'}{4 A B}-\frac{g A' g_E E'}{2 A E_p}+\frac{g^2 B^{''}}{2 B}+\frac{g^2 B' C'}{4 B C}-\frac{g^2 B' f_E E'}{2 B f E_p}-\frac{g B' g_E E'}{B E_p}-\frac{g^2 B^{'2}}{4 B^2}-\\&& \nonumber
\frac{g C' g_E E'}{2 C E_p}+\frac{g f_E g_E E^{'2}}{f E_p^2}-\frac{g g_{EE} E^{'2}}{E_p^2}-\frac{g g_E E^{''}}{E_p}+\frac{2 g^{'2} E^{'2}}{E_p^2},\\
R_{3}^{3}&=&\frac{g^2 A' C'}{4 A C}-\frac{g A' g_E E'}{2 A E_p}+\frac{g^2 B' C'}{4 B C}-\frac{g B' g_E E'}{2 B E_p}+\frac{g^2 C^{''}}{2 C}-\frac{g^2 C' f_E E'}{2 C f E_p}-\frac{g C' g_E E'}{C E_p}-\\&& \nonumber
\frac{g^2 C^{'2}}{4 C^2}+\frac{g f_E g_E E^{'2}}{f E_p^2}-\frac{g g_{EE} E^{'2}}{E_p^2}-\frac{g g_E E^{''}}{E_p}+\frac{2 g_E^2 E^{'2}}{E_p^2}.
\end{eqnarray}
The Ricci scalar is defined by
\begin{eqnarray}
R&=&g^{ij}R_{ij}
\end{eqnarray}
We utilize above definition and get the following expression for Ricci scalar:
\begin{eqnarray}
R&=&\frac{g^2 A^{''}}{A}+\frac{g^2 A' B'}{2 A B}+\frac{g^2 A' C'}{2 A C}-\frac{2 g^2 A' f_E E'}{A f E_p}-\frac{g A' g_E E'}{A E_p}-\frac{g^2 A^{'2}}{2 A^2}+\frac{g^2 B^{''}}{B}+\\&& \nonumber
\frac{g^2 B' C'}{2 B C}-\frac{g^2 B' f_E E'}{B f E_p}-\frac{2 g B' g_E E'}{B E_p}-\frac{g^2 B^{'2}}{2 B^2}+\frac{g^2 C^{''}}{C}-\frac{g^2 C' f_E E'}{C f E_p}-\frac{2 g C' g_E E'}{C E_p}-\\&& \nonumber
\frac{g^2 C^{'2}}{2 C^2}-\frac{2 g^2 f_{EE} E^{'2}}{f E_p^2}-\frac{2 g^2 f_E E^{''}}{f E_p}+\frac{2 g f_E g_E E^{'2}}{f E_p^2}+\frac{4 g^2 f_E^2 E^{'2}}{f^2 E_p^2}-\frac{4 g g_{EE} E^{'2}}{E_p^2}-\\&& \nonumber
\frac{4 g g_E E^{''}}{E_p}+\frac{6 g_E^2 E^{'2}}{E_p^2}.
\end{eqnarray}
 The covariant derivative for a vector $B_\mu$ is defined by
\begin{eqnarray}
\nabla_\mu B_\nu&=&\partial_\mu B_\nu-\Gamma^\alpha_{\mu\nu} B_\alpha.
\end{eqnarray}
From the above definition, we can compute the components of $\nabla_\mu\nabla_\mu F$
\begin{eqnarray}
\nabla_t\nabla_t F&=&\ddot{F}-\Gamma^r_{tt} F^{'},\\
\nabla_r\nabla_r F&=&F^{''}-\Gamma^r_{rr} F^{'}=F^{''},\\
\nabla_\varphi\nabla_\varphi F&=&F_{\varphi\varphi}-\Gamma^r_{\varphi\varphi} F^{'}=-\Gamma^r_{\varphi\varphi} F^{'},\\
\nabla_z\nabla_z F&=&F_{zz}-\Gamma^r_{zz} F^{'}=-\Gamma^r_{zz} F^{'}.
\end{eqnarray}
Now, we are able to calculate the all four components of quantity $A_\mu$ defined in (\ref{aa}) as
\begin{eqnarray}
A_t&=& \frac{g^2 F' A'}{2 A}-\frac{f^2 \ddot{F}}{A}-\frac{g^2 F^{'} f_E E'}{f E_p}+\frac{F g^2 A^{''}}{2 A}+\frac{F g^2 A' B'}{4 A B}+\frac{F g^2 A' C'}{4 A C}-\frac{F g^2 A' f_E E'}{A f E_p}\\  \nonumber
&-&\frac{F g A' g_E E'}{2 A E_p}-\frac{F g^2 A^{'2}}{4 A^2}-\frac{F g^2 B' f_E E'}{2 B f E_p}-\frac{F g^2 C' f_E E'}{2 C f E_p}-\frac{F g^2 f_{EE} E^{'}}{f E_p^2}-\frac{F g^2 f_E E^{''}}{f E_p}\\  \nonumber
&+&\frac{F g f_E g_E E^{'2}}{f E_p^2}+\frac{2 F g^2 f_E^2 E^{'2}}{f^2 E_p^2},\\
A_r&=& g^2 F^{''}+\frac{g F^{'} g_E E'}{E_p}+\frac{F g^2 A^{''}}{2 A}-\frac{F g^2 A' f_E E'}{A f E_p}+\frac{F g A' g_E E'}{2 A E_p}-\frac{F g^2 A^{'2}}{4 A^2}+\frac{F g^2 B^{''}}{2 B}\\  \nonumber
&-&\frac{F g B' g_E E'}{2 B E_p}-\frac{F g^2 B^{'2}}{4 B^2}+\frac{F g^2 C^{''}}{2 C}-\frac{F g C' g_E E'}{2 C E_p}-\frac{F g^2 C^{'2}}{4 C^2}-\frac{F g^2 f^{''} E^{'2}}{f E_p^2}-\frac{F g^2 f_E E^{''}}{f E_p}\\  \nonumber
&-&\frac{F g f_E g_E E^{'2}}{f E_p^2}+\frac{2 F g^2 f_E^2 E^{'2}}{f^2 E_p^2}-\frac{2 F g g_{EE} E^{'2}}{E_p^2}- 
\frac{2 F g g_E E^{''}}{E_p}+\frac{2 F g_{EE}^2 E^{'2}}{E_p^2},\\
A_{\varphi} &=& \frac{g^2 F^{'} B'}{2 B}-\frac{g F^{'} g_E E'}{E_p}+\frac{F g^2 A' B'}{4 A B}-\frac{F g A' g_E E'}{2 A E_p}+\frac{F g^2 B^{''}}{2 B}+\frac{F g^2 B' C'}{4 B C}-\frac{F g^2 B' f_E E'}{2 B f E_p}\\  \nonumber
&-&\frac{F g B' g_E E'}{B E_p}-\frac{F g^2 B^{'2}}{4 B^2}-\frac{F g C' g_E E'}{2 C E_p}+\frac{F g f_E g_E E^{'2}}{f E_p^2}- 
\frac{F g g_{EE} E^{'2}}{E_p^2}-\frac{F g g_E E^{''}}{E_p}+\frac{2 F g_{E}^{2} E^{'2}}{E_p^2},\\
A_z&=& \frac{g^2 F^{'} C'}{2 C}-\frac{g F^{'} g_E E'}{E_p}+\frac{F g^2 A' C'}{4 A C}-\frac{F g A' g_E E'}{2 A E_p}+\frac{F g^2 B' C'}{4 B C}-\frac{F g B' g_E E'}{2 B E_p}+\frac{F g^2 C^{''}}{2 C}\\ \nonumber
&-&\frac{F g^2 C' f_E E'}{2 C f E_p}-\frac{F g C' g_E E'}{C E_p}-\frac{F g^2 C^{'2}}{4 C^2}+\frac{F g f_E g_E E^{'2}}{f E_p^2}- 
\frac{F g g_{EE} E^{'2}}{E_p^2}-\frac{F g g_E E^{''}}{E_p}+\frac{2 F g_{E}^2 E^{'2}}{E_p^2}.
\end{eqnarray}
This enables us to write the following independent
field equations:
\begin{eqnarray}
\frac{g^2 F^{'} A'}{2 A}-\frac{f^2 \ddot{F}}{A}-\frac{g^2 F^{'} f_E E'}{f E_p}-g^2 F^{''}-\frac{g F^{'} g_E E'}{E_p}+\frac{F g^2 A' B'}{4 A B}+\frac{F g^2 A' C'}{4 A C}-&&\\ \nonumber
\frac{F g A' g_E E'}{A E_p}-\frac{F g^2 B^{''}}{2 B}-\frac{F g^2 B' f_E E'}{2 B f E_p}+\frac{F g B' g_E E'}{2 B E_p}+\frac{F g^2 B^{'2}}{4 B^2}-\frac{F g^2 C^{''}}{2 C}-\frac{F g^2 C' f_E E'}{2 C f E_p}+&&\\ \nonumber
\frac{F g C' g_E E'}{2 C E_p}+\frac{F g^2 C^{'2}}{4 C^2}+\frac{2 F g f_E g_E E^{'2}}{f E_p^2}+\frac{2 F g g_{EE} E^{'2}}{E_p^2}+\frac{2 F g g_E E^{''}}{E_p}-\frac{2 F g_E^{2} E^{'2}}{E_p^2}&=&0,\\
\frac{g^2 F^{'} A'}{2 A}-\frac{f^2 \ddot{F}}{A}-\frac{g^2 F^{'} B'}{2 B}-\frac{g^2 F^{'} f_E E'}{f E_p}+\frac{g F^{'} g_E E'}{E_p}+\frac{F g^2 A^{''}}{2 A}+\frac{F g^2 A' C'}{4 A C}-\frac{F g^2 A' f_E E'}{A f E_p}-&&\\ \nonumber
\frac{F g^2 A^{'2}}{4 A^2}-\frac{F g^2 B^{''}}{2 B}-\frac{F g^2 B' C'}{4 B C}+\frac{F g B' g_E E'}{B E_p}+\frac{F g^2 B^{'2}}{4 B^2}-\frac{F g^2 C' f_E E'}{2 C f E_p}+\frac{F g C' g_E E'}{2 C E_p}-&&\\ \nonumber
\frac{F g^2 f_{EE} E^{'2}}{f E_p^2}-\frac{F g^2 f_E E^{''}}{f E_p}+\frac{2 F g^2 f_E^2 E^{'2}}{f^2 E_p^2}+\frac{F g g_{EE} E^{'2}}{E_p^2}+\frac{F g g_E E^{''}}{E_p}-\frac{2 F g_E^{2} E^{'2}}{E_p^2}&=&0, \\
\frac{g^2 F^{'} A'}{2 A}-\frac{f^2 \ddot{F}}{A}-\frac{g^2 F^{'} C'}{2 C}-\frac{g^2 F^{'} f_E E'}{f E_p}+\frac{g F^{'} g_E E'}{E_p}+\frac{F g^2 A^{''}}{2 A}+\frac{F g^2 A' B'}{4 A B}-&&\\ \nonumber
\frac{F g^2 A' f_E E'}{A f E_p}-\frac{F g^2 A^{'2}}{4 A^2}-\frac{F g^2 B' C'}{4 B C}-\frac{F g^2 B' f_E E'}{2 B f E_p}+\frac{F g B' g_E E'}{2 B E_p}-\frac{F g^2 C^{''}}{2 C}+\frac{F g C' g_E E'}{C E_p}+&&\\ \nonumber
\frac{F g^2 C^{'2}}{4 C^2}-\frac{F g^2 f_{EE} E^{'2}}{f E_p^2}-\frac{F g^2 f_E E^{''}}{f E_p}+\frac{2 F g^2 f_E^2 E^{'2}}{f^2 E_p^2}+\frac{F g g_{EE} E^{'2}}{E_p^2}+\frac{F g g_E E^{''}}{E_p}-\frac{2 F g_E^{2} E^{'2}}{E_p^2}&=&0.
\end{eqnarray}
corresponding to 
$ A_t = A_r, A_t = A_\varphi$ and $A_t = A_z$ respectively.
\section{Time-dependent cosmic strings} 
In this section, we discuss the time-dependent solutions (cosmic strings) in gravity's rainbow. 
Although many different mass configurations lead to static and time-independent metrics, there are some examples with time-dependent results and, therefore, it is worth studying. In this non-static case,  the most important difference with the static one is the structure of the spacetime, as in the former case there are only two parameters in the metric $g_{\varphi\varphi} \left(E/E_{p}\right)$  reducing  to cosmic strings. One more reason to study the cylindrical solutions with time-dependent  fields could be existence of a challenge between spherically  and cylindrically symmetries. In  GR, according to the Birkhoff theorem, we know that   there always exist   a timelike Killing vector $\partial_t$  in the spherically symmetric vacuum metrics. Consequently, we  easily conclude  that the spherically symmetric vacuum gravitating system is necessarily static, i.e., time independent. However, a   dramatic  change occurs when one considers  the cylindrically symmetric systems because there is no such theorem in case of cylindrical symmetry analogous to  Birkhoff’s theorem. Propagation of gravitational waves during the gravitational collapse of a cylindrically symmetric system could be a reason to study time-dependent cylindrical objects\cite{Delice:2004wk}.  In the gravity's rainbow scenario, if we can find a static solution with time dependent mass (energy) factor $f^{2} \left(E(t)/E_{p}\right),g^{2} \left(E(t)/E_{p}\right)$, then our solution  could be a subset of the most general class of Einstein-Rosen (ER) gravitational wave solutions in gravity's rainbow in comparison with the similar  GR solutions obtained in  \cite{Ozgur1,Delice:2004wk}.
%%%%%%%%%%%%%%%%%%%%%%%

Let us to start our  analysis by writing the field equations for the following time-dependent metric,
\begin{equation}
ds^{2} = \frac{A(t,r)}{f^{2} \left(E/E_{p}\right)} dt^{2} - \frac{1}{g^{2} \left(E/E_{p}\right)}dr^{2} - \frac{B(t,r)}{g^{2} \left(E/E_{p}\right)} d \varphi^{2} - \frac{C(t,r)}{g^{2} \left(E/E_{p}\right)} dz^{2}.\label{metric2}
\end{equation}
By implementing metric (\ref{metric2}) to (\ref{feq}), we obtain the following set of field equations:
\begin{eqnarray}
&& \frac{g^2}{4}\left[\left(\frac{B^{'}}{B}\right)^2+\left(\frac{C^{'}}{C}\right)^2-\frac{B^{'}C^{'}}{BC}-2\frac{B^{''}}{B}-2\frac{C^{''}}{C}\right]+\frac{1}{4}\frac{\dot{B}\dot{C}}{BC} = \rho,\\
&&\frac{A^{'}\dot{B}}{AB}+\frac{A^{'}\dot{C}}{AC}-2\frac{\dot{B}^{'}}{B}+\frac{B^{'}\dot{B}}{B^2}-2\frac{\dot{C}^{'}}{C}+\frac{C^{'}\dot{C}}{C^2} = 0,\\
&&\frac{1}{4}\left(\frac{A^{'} B^{'}}{AB}+\frac{1}{4}\frac{A^{'}C^{'}}{AC}+\frac{1}{4}\frac{B^{'}C^{'}}{BC}\right)+ 
 \frac{1}{4}\frac{f^2}{A}\left[\frac{\dot{A}\dot{B}}{AB}+\frac{\dot{A}\dot{C}}{AC}-\frac{\dot{B}\dot{C}}{BC}+\left(\frac{\dot{B}}{B}\right)^2+\left(\frac{\dot{C}}{C}\right)^2
 \right.\\
 &&\left. -2\frac{\ddot{B}}{B}-2\frac{\ddot{C}}{C}\right]=p_r,\\
&&\frac{1}{4}\left[\frac{A^{'}C^{'}}{AC}+2\frac{A^{''}}{A}+2\frac{C^{''}}{C}-\left(\frac{A^{'}}{A}\right)^2-\left(\frac{C^{'}}{C}\right)^2\right]+\frac{1}{4}\frac{f^2}{Ag^2}\left[\frac{\dot{A}\dot{C}}{AC}-2\frac{\ddot{C}}{C}+\left(\frac{\dot{C}}{C}\right)^2\right]=\frac{p_\varphi}{g^2},\\
&&\frac{1}{4}\left[\frac{A^{'}B^{'}}{AB}+2\frac{A^{''}}{A}+2\frac{B^{''}}{B}-\left(\frac{A^{'}}{A}\right)^2-\left(\frac{B^{'}}{B}\right)^2\right]+\frac{1}{4}\frac{f^2}{Ag^2}\left[\frac{\dot{A}\dot{B}}{AB}-2\frac{\ddot{B}}{B}+\left(\frac{\dot{B}}{B}\right)^2\right]=\frac{p_z}{g^2}.
\end{eqnarray}
Here ``prime" and ``dot" indicate  the derivative with respect to $r$ and $ t$ respectively. A   solution can possibly  be obtained in vacuum when $T_{\mu}^{\nu}=0$ with the assumption   that all metric functions depend on time  only, this means that all primed functions will vanish. With these assumptions, we can show that the second field equation  is satisfied identically and other equations reduce to have  following form:
\begin{eqnarray}
&&\frac{\dot{B}\dot{C}}{BC}=0,\\
&&\frac{\dot{A}\dot{B}}{AB}+\frac{\dot{A}\dot{C}}{AC}-\frac{\dot{B}\dot{C}}{BC}+\left(\frac{\dot{B}}{B}\right)^2+\left(\frac{\dot{C}}{C}\right)^2-2\frac{\ddot{B}}{B}-2\frac{\ddot{C}}{C}=0,\\
&&\frac{\dot{A}\dot{C}}{AC}-2\frac{\ddot{C}}{C}+\left(\frac{\dot{C}}{C}\right)^2=0\\
&&\frac{\dot{A}\dot{B}}{AB}-2\frac{\ddot{B}}{B}+\left(\frac{\dot{B}}{B}\right)^2=0.
\end{eqnarray}
Here, we found three class of exact solutions for time-dependent gravity's rainbow. 
\begin{itemize}
\item The first class  of exact solutions is given by
\begin{equation}
ds^{2} = \frac{A(t)}{f^{2} \left(E/E_{p}\right)} dt^{2} - \frac{1}{g^{2} \left(E/E_{p}\right)}dr^{2} - \frac{B_0}{g^{2} \left(E/E_{p}\right)} d \varphi^{2} - \frac{C_0}{g^{2} \left(E/E_{p}\right)} dz^{2},
\end{equation}
where $A(t)$ is an arbitrary function of time and $B_0,C_0$ are arbitrary constants. 
\item The second family of exact solutions is as following:
\begin{equation}
ds^{2} = \frac{c_1\dot{C}^2}{C(t)f^{2} \left(E/E_{p}\right)} dt^{2} - \frac{1}{g^{2} \left(E/E_{p}\right)}dr^{2} - \frac{B(t)}{g^{2} \left(E/E_{p}\right)} d \varphi^{2} - \frac{C(t)}{g^{2} \left(E/E_{p}\right)} dz^{2},
\end{equation}
where $C(t),B(t)$ are arbitrary time functions and $c_1$  is a constant. 
\item The last member of exact solutions is given by the following metric:
\begin{equation}
ds^{2} = \frac{c_1\dot{B}^2}{B(t)f^{2} \left(E/E_{p}\right)} dt^{2} - \frac{1}{g^{2} \left(E/E_{p}\right)}dr^{2} - \frac{B(t)}{g^{2} \left(E/E_{p}\right)} d \varphi^{2} - \frac{C_0}{g^{2} \left(E/E_{p}\right)} dz^{2}.
\end{equation}
\end{itemize}
Importantly,  we  note that here $E=E(t)$ and metrics are completely different from any other GR solutions. 
%%%%%%%%%%%%%%%%%%%%%%%%%%%%%%
\section{When $E=E(t)$}
The line element in this case of gravity's rainbow is given by
\begin{equation}
ds^{2} = \frac{A(t,r)}{f^{2} \left(E(t)/E_{p}\right)} dt^{2} - \frac{1}{g^{2} \left(E(t)/E_{p}\right)}dr^{2} - \frac{B(t,r)}{g^{2} \left(E(t)/E_{p}\right)} d \varphi^{2} - \frac{C(t,r)}{g^{2} \left(E(t)/E_{p}\right)} dz^{2}.\label{metric4}
\end{equation}
It is clearly evident that
 the   expressions of the nonzero components of the metric tensor are
$$
g_{00} = \frac{A(t,r)}{f^{2} \left(E(t)/E_{p}\right)} , \  g_{11} = - \frac{1}{g^{2} \left(E(r)/E_{p}\right)}, \  g_{22} = - \frac{B(t,r)}{g^{2} \left(E(t)/E_{p}\right)}, \  g_{33} = - \frac{C(t,r)}{g^{2} \left(E(t)/E_{p}\right)}.
$$
The inverse metric components are, 
$$
g^{00} = \frac{f^{2} \left(E(t)/E_{p}\right)}{A(t,r)}, \  g^{11}= - g^{2} \left(E(t)/E_{p}\right), \ g^{22} = - \frac{g^{2} \left(E(t)/E_{p}\right)}{B(t,r)}, \  g^{33} = - {\frac{g^{2} \left(E(t)/E_{p}\right)}{C(t,r)}}.
$$
With these values of metric components, 
 it is easy to calculate the  Christoffel symbols of the second kind  
\begin{equation}
\Gamma_{jl}^{i} = \frac{1}{2} g^{im} \left(\frac{\partial g_{mj}}{\partial x^{l}} + \frac{\partial g_{ml}}{\partial x^{j}} - \frac{\partial g_{jl}}{\partial x^{m}}\right).
\end{equation}
The calculation leads to the following expressions: 
\begin{eqnarray}
\Gamma^{0}_{00}&=&\frac{\dot{A}}{2 A}-\frac{f_E \dot{E}}{f E_p},\ \Gamma^{1}_{10}=\frac{A^{'}}{2 A}, \ \Gamma^{0}_{11}=-\frac{f^2 g_E \dot{E}}{A g^3 E_p}, \ \Gamma^{1}_{22}=\frac{f^2 \dot{B}}{2 A g^2}-\frac{B f^2 g_E \dot{E}}{A g^3 E_p},\\ \nonumber
\Gamma^{1}_{33}&=&\frac{f^2 \dot{C}}{2 A g^2}-\frac{C f^2 g_E \dot{E}}{A g^3 E_p},\ \Gamma^{1}_{00}=\frac{g^2 A^{'}}{2 f^2},\ \Gamma^{1}_{10}=-\frac{g_E \dot{E}}{g E_p}, \ \Gamma^{1}_{22}=-\frac{1}{2} B^{'},\\ \nonumber
 \ \Gamma^{1}_{33}&=&-\frac{1}{2} C^{'}, \ \Gamma^{2}_{20}=\frac{\dot{B}}{2 B}-\frac{g_E \dot{E}}{g E_p}, \ \Gamma^{2}_{21}=\frac{B^{'}}{2 B}, \ \Gamma^{3}_{30}=\frac{\dot{C}}{2 C}-\frac{g_E \dot{E}}{g E_p}, \ \Gamma^{3}_{31}=\frac{C^{'}}{2 C}.
\end{eqnarray}
 Exploiting definition of Ricci tensor (\ref{ri}), the components are calculated by
\begin{eqnarray}
R_{00}&=&\frac{g^2 A^{'} B^{'}}{4 B f^2}+\frac{\dot{A} \dot{B}}{4 A B}+\frac{g^2 A^{'} C^{'}}{4 C f^2}+\frac{\dot{A} \dot{C}}{4 A C}+\frac{g^2 A^{''}}{2 f^2}-\frac{g^2 A^{'2}}{4 A f^2}-\frac{3 \dot{A} g_E \dot{E}}{2 A g E_p}-\frac{\dot{B} f_E \dot{E}}{2 B f E_p}+\\ &&\nonumber
\frac{\dot{B} g_E \dot{E}}{B g E_p}-\frac{\ddot{B}}{2 B}+\frac{\dot{B}^2}{4 B^2}-\frac{\dot{C} f_E \dot{E}}{2 C f E_p}+\frac{\dot{C} g_E \dot{E}}{C g E_p}-\frac{\ddot{C}}{2 C}+\frac{\dot{C}^2}{4 C^2}+\frac{3 f_E g_E \dot{E}^2}{f g E_p^2}+\\ &&\nonumber
\frac{3 g_{EE} \dot{E}^2}{g E_p^2}+\frac{3 g_E \ddot{E}}{g E_p}-\frac{6 g_E^2 \dot{E}^2}{g^2 E_p^2}, \\
R_{10}&=&\frac{A^{'} B^{'}}{4 A B}+\frac{A^{'} \dot{C}}{4 A C}-\frac{A^{'} g_E \dot{E}}{A g E_p}-\frac{\dot{B}^{'}}{2 B}+\frac{B^{'} \dot{B}}{4 B^2}-\frac{\dot{C}^{'}}{2 C}+\frac{C^{'} \dot{C}}{4 C^2},\\
R_{11}&=&-\frac{A^{''}}{2 A}+\frac{f^2 \dot{A} g_E \dot{E}}{2 A^2 g^3 E_p}+\frac{A^{'2}}{4 A^2}-\frac{f^2 \dot{B} g_E \dot{E}}{2 A B g^3 E_p}-\frac{f^2 \dot{C} g_E \dot{E}}{2 A C g^3 E_p}-\frac{B^{''}}{2 B}+\\ &&\nonumber
\frac{B^{'2}}{4 B^2}-\frac{C^{''}}{2 C}+\frac{C^{'2}}{4 C^2}+\frac{4 f^2 g_E^2 \dot{E}^2}{A g^4 E_p^2}-\frac{f^2 g_{EE} \dot{E}^2}{A g^3 E_p^2}-\frac{f^2 g_E \dot{E}}{A g^3 E_p}-\frac{f f_E g_E \dot{E}^2}{A g^3 E_p^2},\\
R_{22}&=&-\frac{A^{'} B^{'}}{4 A}-\frac{f^2 \dot{A} \dot{B}}{4 A^2 g^2}+\frac{B f^2 \dot{A} g_E \dot{E}}{2 A^2 g^3 E_p}+\frac{f^2 \dot{B} \dot{C}}{4 A C g^2}+\frac{f^2 \ddot{B}}{2 A g^2}-\frac{f^2 \dot{B}^2}{4 A B g^2}-\frac{2 f^2 \dot{B} g_E \dot{E}}{A g^3 E_p}+\\ &&\nonumber
\frac{f \dot{B} f_E \dot{E}}{2 A g^2 E_p}-\frac{B f^2 \dot{C} g_E \dot{E}}{2 A C g^3 E_p}-\frac{B^{'} C^{'}}{4 C}+\frac{B^{'2}}{4 B}-\frac{1}{2} B^{''}+\frac{4 B f^2 g_E^2 \dot{E}}{A g^4 E_p^2}-\frac{B f^2 g_{EE} \dot{E}^2}{A g^3 E_p^2}-\\ &&\nonumber
\frac{B f^2 g_E \ddot{E}}{A g^3 E_p}-\frac{B f f_E g_E \dot{E}^2}{A g^3 E_p^2},\\
R_{33}&=&-\frac{A^{'} C^{'}}{4 A}-\frac{f^2 \dot{A} \dot{C}}{4 A^2 g^2}+\frac{C f^2 \dot{A} g_E \dot{E}}{2 A^2 g^3 E_p}+\frac{f^2 \dot{B} \dot{C}}{4 A B g^2}-\frac{C f^2 \dot{B} g_E \dot{E}}{2 A B g^3 E_p}+\frac{f^2 \ddot{C}}{2 A g^2}-\frac{f^2 \dot{C}^2}{4 A C g^2}-\\ &&\nonumber
\frac{2 f^2 \dot{C} g_E \dot{E}}{A g^3 E_p}+\frac{f \dot{C} f_E \dot{E}}{2 A g^2 E_p}-\frac{B^{'} C^{'}}{4 B}+\frac{C^{'2}}{4 C}-\frac{1}{2} C^{''}+\frac{4 C f^2 g_E^2 \dot{E}^2}{A g^4 E_p^2}-\frac{C f^2 g_{EE} \dot{E}^2}{A g^3 E_p^2}-\\ &&\nonumber
\frac{C f^2 g_E \ddot{E}}{A g^3 E_p}-\frac{C f f_E g_E \dot{E}^2}{A g^3 E_p^2}.
\end{eqnarray}
 Now, we compute the expression for Ricci scalar $R=g^{\mu\nu}R_{\mu\nu}$
 as following:
\begin{eqnarray}
R&=&\frac{g^2 A^{'} B^{'}}{2 A B}+\frac{g^2 A^{'} C^{'}}{2 A C}+\frac{g^2 A^{''}}{A}+\frac{f^2 \dot{A} \dot{B}}{2 A^2 B}+\frac{f^2 \dot{A} \dot{C}}{2 A^2 C}-\frac{3 f^2 \dot{A} g_E \dot{E}}{A^2 g E_p}-\frac{g^2 A^{'2}}{2 A^2}-\frac{f^2 \dot{B} \dot{C}}{2 A B C}+\\ &&\nonumber
\frac{4 f^2 \dot{B} g_E \dot{E}}{A B g E_p}-\frac{f^2 \ddot{B}}{A B}-\frac{f \dot{B} f_E \dot{E}}{A B E_p}+\frac{f^2 \dot{B}^2}{2 A B^2}+\frac{4 f^2 \dot{C} g_E \dot{E}}{A C g E_p}-\frac{f^2 \ddot{C}}{A C}-\frac{f \dot{C} f_E \dot{E}}{A C E_p}+\frac{f^2 \dot{C}^2}{2 A C^2}+\\ &&\nonumber
\frac{g^2 B^{'} C^{'}}{2 B C}+\frac{g^2 B^{''}}{B}-\frac{g^2 B^{'2}}{2 B^2}+\frac{g^2 C^{''}}{C}-\frac{g^2 C^{'2}}{2 C^2}+\frac{6 f^2 g_{EE} \dot{E}^2}{A g E_p^2}+\frac{6 f^2 g_E \ddot{E}}{A g E_p}-\\ &&\nonumber
\frac{18 f^2 g_E^2 \dot{E}^2}{A g^2 E_p^2}+\frac{6 f f_E g_E \dot{E}^2}{A g E_p^2}.
\end{eqnarray}
 Now, following section V, the independent
field equation corresponding to $ A_t = A_r$ is 
\begin{eqnarray}
\frac{4 F g_E^2 \dot{E}^2 f^2}{A g^6 E_p^2}+\frac{F \dot{B}^2 f^2}{4 A B^2}+\frac{F \dot{C}^2 f^2}{4 A C^2}+\frac{3 F \dot{E}^2 g_{EE} f^2}{A g E_p^2}+\frac{3 F g_E \ddot{E} f^2}{A g E_p}+\frac{F g_E \dot{E} \dot{A} f^2}{2 A^2 g^5 E_p}+\\ \nonumber
\frac{F g_E \dot{E} \dot{B} f^2}{A B g E_p}+\frac{F \dot{A} \dot{B} f^2}{4 A^2 B}+\frac{F g_E \dot{E} \dot{C} f^2}{A C g E_p}+\frac{F \dot{A} \dot{C} f^2}{4 A^2 C}+\frac{\dot{A} \dot{F} f^2}{2 A^2}-\frac{\ddot{F} f^2}{A}-\frac{F \ddot{B} f^2}{2 A B}-\\ \nonumber
\frac{F \ddot{C} f^2}{2 A C}-\frac{3 F g_E \dot{E} \dot{A} f^2}{2 A^2 g E_p}-\frac{F g_E \ddot{E} f^2}{A g^5 E_p}-\frac{F g_E \dot{E} \dot{B} f^2}{2 A B g^5 E_p}-\frac{F g_E \dot{E} \dot{C} f^2}{2 A C g^5 E_p}-\\ \nonumber
\frac{6 F g_E^2 \dot{E}^2 f^2}{A g^2 E_p^2}-\frac{F \dot{E}^2 g_{EE} f^2}{A g^5 E_p^2}-\frac{g_E \dot{E} \dot{F} f^2}{A g^5 E_p^2}+\frac{3 F f_E g_E \dot{E}^2 f}{A g E_p^2}-\frac{F f_E \dot{E} \dot{B} f}{2 A B E_p}-\\ \nonumber
\frac{F f_E \dot{E} \dot{C} f}{2 A C E_p}-\frac{F f_E g_E \dot{E}^2 f}{A g^5 E_p^2}+\frac{F A^{'2}}{4 A^2 g^2}+\frac{F B^{'2}}{4 B^2 g^2}+\frac{F C^{'2}}{4 C^2 g^2}+\frac{F g^2 A^{'} B^{'}}{4 A B}+\frac{F g^2 A^{'} C^{'}}{4 A C}+\\ \nonumber
\frac{g^2 A^{'} F^{'}}{2 A}+\frac{F g^2 A^{''}}{2 A}-\frac{F g^2 A^{'2}}{4 A^2}-\frac{F^{''}}{g^2}-\frac{F A^{''}}{2 A g^2}-\frac{F B^{''}}{2 B g^2}-\frac{F C^{''}}{2 C g^2}&=&0,
\end{eqnarray}
The independent equations corresponding to  $A_t = A_\varphi$ and $A_t = A_z$ are
given  respectively by,
\begin{eqnarray}
\frac{4 F g_E^2 \dot{E}^2 f^2}{A g^6 E_p^2}+\frac{F \dot{B}^2 f^2}{4 A B^2}+\frac{F \dot{C}^2 f^2}{4 A C^2}+\frac{3 F \dot{E}^2 g_{EE} f^2}{A g E_p^2}+\frac{3 F g_E \ddot{E} f^2}{A g E_p}+\frac{F g_E \dot{E} \dot{A} f^2}{2 A^2 g^5 E_p}+\\ \nonumber
\frac{F g_E \dot{E} \dot{B} f^2}{A B g E_p}+\frac{F^{'} \dot{B} f^2}{2 A B g^4}+\frac{F \dot{A} \dot{B} f^2}{4 A^2 B}+\frac{F g_E \dot{E} \dot{C} f^2}{A C g E_p}+\frac{F \dot{A} \dot{C} f^2}{4 A^2 C}+\frac{F \dot{B} \dot{C} f^2}{4 A B C g^4}+\frac{\dot{A} \dot{F} f^2}{2 A^2}+\\ \nonumber
\frac{F \ddot{B} f^2}{2 A B g^4}-\frac{\ddot{F} f^2}{A}-\frac{F \ddot{B} f^2}{2 A B}-\frac{F \ddot{C} f^2}{2 A C}-\frac{F \dot{A} \dot{B} f^2}{4 A^2 B g^4}-\frac{F \dot{B}^2 f^2}{4 A B^2 g^4}-\frac{3 F g_E \dot{E} \dot{A} f^2}{2 A^2 g E_p}-\frac{F g_E \ddot{E} f^2}{A g^5 E_p}-\\ \nonumber
\frac{2 F g_E \dot{E} \dot{B} f^2}{A B g^5 E_p}-\frac{F g_E \dot{E} \dot{C} f^2}{2 A C g^5 E_p}-\frac{6 F g_E^2 \dot{E}^2 f^2}{A g^2 E_p^2}-\frac{F \dot{E}^2 g_{EE} f^2}{A g^5 E_p^2}+\frac{3 F f_E g_E \dot{E}^2 f}{A g E_p^2}+\frac{F f_E \dot{E} \dot{B} f}{2 A B g^4 E_p}-\\ \nonumber
\frac{F f_E \dot{E} \dot{B} f}{2 A B E_p}-\frac{F f_E \dot{E} \dot{C} f}{2 A C E_p}-\frac{F f_E g_E \dot{E}^2 f}{A g^5 E_p^2}+\frac{F B^{'2}}{4 B^2 g^2}+\frac{F g^2 A^{'} B^{'}}{4 A B}+\frac{F g^2 A^{'} C^{'}}{4 A C}+\\ \nonumber
\frac{g^2 A^{'} F^{'}}{2 A}+\frac{F g^2 A^{''}}{2 A}-\frac{F g^2 A^{'2}}{4 A^2}-\frac{F B^{''}}{2 B g^2}-\frac{F A^{'} B^{'}}{4 A B g^2}-\frac{F B^{'} C^{'}}{4 B C g^2}&=&0, \\
\frac{4 F g_E^2 \dot{E}^2 f^2}{A g^6 E_p^2}+\frac{F \dot{B}^2 f^2}{4 A B^2}+\frac{F \dot{C}^2 f^2}{4 A C^2}+\frac{3 F \dot{E}^2 g_{EE} f^2}{A g E_p^2}+\frac{3 F g_E \ddot{E} f^2}{A g E_p}+\frac{F g_E \dot{E} \dot{A} f^2}{2 A^2 g^5 E_p}+\\ \nonumber
\frac{F g_E \dot{E} \dot{B} f^2}{A B g E_p}+\frac{F \dot{A} \dot{B} f^2}{4 A^2 B}+\frac{F g_E \dot{E} \dot{C} f^2}{A C g E_p}+\frac{F \dot{A} \dot{C} f^2}{4 A^2 C}+\frac{F \dot{B} \dot{C} f^2}{4 A B C g^4}+\frac{\dot{A} \dot{F} f^2}{2 A^2}+\frac{F \ddot{C} f^2}{2 A C g^4}-\\ \nonumber
\frac{\ddot{F} f^2}{A}-\frac{F \ddot{B} f^2}{2 A B}-\frac{F \ddot{C} f^2}{2 A C}-\frac{F \dot{A} \dot{C} f^2}{4 A^2 C g^4}-\frac{F \dot{C}^2 f^2}{4 A C^2 g^4}-\frac{3 F g_E \dot{E} \dot{A} f^2}{2 A^2 g E_p}-\frac{F g_E \ddot{E} f^2}{A g^5 E_p}-\\ \nonumber
\frac{F g_E \dot{E} \dot{B} f^2}{2 A B g^5 E_p}-\frac{2 F g_E \dot{E} \dot{C} f^2}{A C g^5 E_p}-\frac{6 F g_E^2 \dot{E}^2 f^2}{A g^2 E_p^2}-\frac{F \dot{E}^2 g_{EE} f^2}{A g^5 E_p^2}+\frac{3 F f_E g_E \dot{E}^2 f}{A g E_p^2}+\\ \nonumber
\frac{F f_E \dot{E} \dot{C} f}{2 A C g^4 E_p}-\frac{F f_E \dot{E} \dot{B} f}{2 A B E_p}-\frac{F f_E \dot{E} \dot{C} f}{2 A C E_p}-\frac{F f_E g_E \dot{E}^2 f}{A g^5 E_p^2}+\frac{F \dot{C}^2}{4 C^2 g^2}+\frac{F g^2 A^{'} B^{'}}{4 A B}+\\ \nonumber
\frac{F g^2 A^{'} C^{'}}{4 A C}+\frac{g^2 A^{'} F^{'}}{2 A}+\frac{F g^2 A^{''}}{2 A}-\frac{F g^2 A^{'2}}{4 A^2}-\frac{C^{'} F^{'}}{2 C g^2}-\frac{F C^{''}}{2 C g^2}-\frac{F A^{'} C^{'}}{4 A C g^2}-\frac{F B^{'} C^{'}}{4 B C g^2}&=&0.
\end{eqnarray}
Here, any set of functions
 satisfying the above equations would be a solution of the modified
Einstein field equations   for a given $F(r)$ in gravity's rainbow.  
Here, we see that  to find    a general solution to the above equations are not  an easy task.
%%%%%%%%%%%%%%%%%%%%%%%%%%%%%%
\section{Concluding remarks}
The exact solutions play a central role in gravity theory.
However, a deformed formalism 
of special relativity, which modifies the standard dispersion relations  in the order of 
Planck length (commonly known as DSR), is generalized to curved-spacetime.
This generalization is known  as gravity's rainbow and  has found lots of attention recent 
days. Keeping these points in mind, in this work,
 we have investigated the static cylindrical solutions for
Einstein's field equations  in gravity's rainbow. 
The cosmic string metric is supposed to be static (i.e with vanishing off
diagonal components and time independent) and cylindrically symmetric.
In this setting, we have discussed cosmic strings in energy-dependent background.
The fields equations following this metric lead to various energy-dependent 
 differential equations. In order to solve these  differential equations,
 we have considered the possibility of Kasner's, quasi-Kasner and non-Kasner solutions.
It is well-known that the Kasner solutions  are
two parametric metric and  unique exact solutions for the Einstein equations
with cylindrical symmetry. It is shown that  the quasi-Kasner solutions can
not be realized in gravity's rainbow. Also, we
have found that the gravity's rainbow cosmic strings follow same behavior (singularities) 
 to that of standard GR theory. We also analysed the time-dependent solutions (cosmic strings) in gravity's rainbow. Here, to discuss the time-dependent solutions,  we have assumed the vanishing  energy-momentum tensor together with only time-dependent metric functions. Here, we have observed that the metric structures are  completely different to that of the other GR solutions.

%%%%%%%%%%%%%%%%%%%%%%%%%%%%%%%%%%%%%%%%%%%%%%%%%%%%%%%%%%%%%%%%%%%%

\appendix 
\section*{Appendix: Mathematical details}
In this appendix, we present explicit forms of different geometrical quantities which are used to derive field equations.
%%%%%%%%%%%%%%%%%%%%%%%%%%%%%%%%%%%%%%%%%%%%%%%%%%%%%%%%%%%%%%%%%%%%

\subsection{Case $A=A(r)$, $B=B(r)$ and $C=C(r)$}
The  nonzero components of the metric tensor are as following:
$$
g_{00} = \frac{A(r)}{f^{2} \left(E/E_{p}\right)} , \  g_{11} = - \frac{1}{g^{2} \left(E/E_{p}\right)}, \  g_{22} = - \frac{B(r)}{g^{2} \left(E/E_{p}\right)}, \  g_{33} = - \frac{C(r)}{g^{2} \left(E/E_{p}\right)}.
$$

The inverse of these   metric components are given by
$$
g^{00} = \frac{f^{2} \left(E/E_{p}\right)}{A(r)}, \  g^{11}= - g^{2} \left(E/E_{p}\right), \ g^{22} = - \frac{g^{2} \left(E/E_{p}\right)}{B(r)}, \  g^{33} = - {\frac{g^{2} \left(E/E_{p}\right)}{C(r)}}.
$$

Utilizing definition (\ref{chr}), the nonzero values of the Christoffel symbols  are as follows, 
\begin{eqnarray}
\Gamma^{0}_{10}&=&\frac{1}{2} \frac{A^{'}}{A},\ \Gamma^{1}_{00}=\frac{1}{2} \frac{A^{'}}{f^2}g^2,\ \Gamma^{1}_{22}=-\frac{1}{2} B^{'},\\ \nonumber
\Gamma^{1}_{33}&=&-\frac{1}{2} C^{'},\ \Gamma^{2}_{21}=\frac{1}{2} \frac{B^{'}}{B},\ \Gamma^{3}_{31}=\frac{1}{2} \frac{C^{'}}{C}.
\end{eqnarray}

Exploiting the definition (\ref{ri}),  the covariant components of Ricci tensor are calculated as
\begin{eqnarray}
R_{00}&=&-\frac{1}{4}\frac{g^2}{f^2}\frac{A^{'2}}{A}+\frac{1}{4}\frac{g^2}{f^2}\frac{A^{'}B^{'}}{B}+\frac{1}{4}\frac{g^2}{f^2}\frac{A^{'}C^{'}}{C}+\frac{1}{2}\frac{g^2}{f^2}A^{''}, \\
R_{11}&=&\frac{1}{4}\frac{A^{'2}}{A^2}+\frac{1}{4}\frac{B^{'2}}{B^2}+\frac{1}{4}\frac{C^{'2}}{C^2}-\frac{1}{2}\frac{A^{''}}{A}-\frac{1}{2}\frac{B^{''}}{B}-\frac{1}{2}\frac{C^{''}}{C},\\
R_{22}&=&-\frac{1}{4}\frac{A^{'}B^{'}}{A}+\frac{1}{4}\frac{B^{'2}}{B}-\frac{1}{4}\frac{B^{'}C^{'}}{C}-\frac{1}{2}B^{''},\\
R_{33}&=&-\frac{1}{4}\frac{A^{'}C^{'}}{A}-\frac{1}{4}\frac{B^{'}C^{'}}{C}+\frac{1}{4}\frac{C^{'2}}{C}-\frac{1}{2}C^{''},
\end{eqnarray}
and similarly the mixed components are computed as
\begin{eqnarray}
R_{0}^{0}&=&-\frac{g^2}{4} \frac{A^{'2}}{A^2}+\frac{g^2}{4} \left(\frac{A^{'}B^{'}}{AB}\right)^2+\frac{g^2}{4} \left(\frac{A^{'}C^{'}}{AC}\right)^2+\frac{g^2}{2} \frac{A^{''}}{A}, \\
R_{1}^{1}&=&-\frac{g^2}{4}\frac{A^{'2}}{A^2}-\frac{g^2}{4}\frac{B^{'2}}{B^2}-\frac{g^2}{4}\frac{C^{'2}}{C^2}+\frac{g^2}{2} \frac{A^{''}}{A}+\frac{g^2}{2} \frac{B^{''}}{B}+\frac{g^2}{2} \frac{C^{''}}{C},\\
R_{2}^{2}&=&\frac{g^2}{4}\frac{A^{'}B^{'}}{AB}-\frac{g^2}{4}\frac{B^{'2}}{B^2}+\frac{g^2}{4}\frac{B^{'}C^{'}}{BC}+\frac{g^2}{2} \frac{B^{''}}{B},\\
R_{3}^{3}&=&\frac{g^2}{4}\frac{A^{'}C^{'}}{AC}-\frac{g^2}{4}\frac{C^{'2}}{C^2}+\frac{g^2}{4}\frac{B^{'}C^{'}}{BC}+\frac{g^2}{2} \frac{C^{''}}{C},
\end{eqnarray}
Finally, using definition $R = g^{\mu\nu}R_{\mu\nu}$, the expression for Ricci scalar is given by,
 \begin{eqnarray}
R&=&-\frac{g^2}{2} \frac{A^{'2}}{A^2}+\frac{g^2}{2}\frac{A^{'}B^{'}}{AB}-\frac{g^2}{2}\frac{B^{'2}}{B^2}+\frac{g^2}{2}\frac{A^{'}C^{'}}{AC}+\frac{g^2}{2}\frac{B^{'}C^{'}}{BC}-\frac{g^2}{2} \frac{C^{'2}}{C^2}+g^2\frac{A^{''}}{A}+g^2\frac{B^{''}}{B}+g^2\frac{C^{''}}{C}.
\end{eqnarray}

%%%%%%%%%%%%%%%%%%%%%%%%%%%%%%%%%%%%%%%%%%%%%%%%%%%%%%%%%%%%%%%%%%%%

\subsection{Case $A=A(t,r)$, $B=B(t,r)$ and $C=C(t,r)$}
The nonzero components of time-dependent   metric tensor are
$$
g_{00} = \frac{A(t,r)}{f^{2} \left(E/E_{p}\right)} , \  g_{11} = - \frac{1}{g^{2} \left(E/E_{p}\right)}, \  g_{22} = - \frac{B(t,r)}{g^{2} \left(E/E_{p}\right)}, \  g_{33} = - \frac{C(t,r)}{g^{2} \left(E/E_{p}\right)}.
$$

The inverse of these metric components are
$$
g^{00} = \frac{f^{2} \left(E/E_{p}\right)}{A(t,r)}, \  g^{11}= - g^{2} \left(E/E_{p}\right), \ g^{22} = - \frac{g^{2} \left(E/E_{p}\right)}{B(t,r)}, \  g^{33} = - {\frac{g^{2} \left(E/E_{p}\right)}{C(t,r)}}.
$$

The Christoffel symbols of the second kind are,
\begin{eqnarray}
\Gamma^{0}_{00}&=&\frac{1}{2} \frac{\dot{A}}{A},\ \Gamma^{0}_{10}=\frac{1}{2} \frac{A^{'}}{A},\ \Gamma^{0}_{22}=\frac{1}{2}\frac{f^2}{g^2}\frac{\dot{B}}{A},\ \Gamma^{0}_{33}=\frac{1}{2}\frac{f^2}{g^2}\frac{\dot{C}}{A},\ \Gamma^{1}_{00}=\frac{1}{2}\frac{g^2}{f^2}A^{'},\\ \nonumber \Gamma^{1}_{22}&=&-\frac{1}{2} B^{'},\ \Gamma^{1}_{33}=-\frac{1}{2} C^{'},\ \Gamma^{2}_{20}=\frac{1}{2}\frac{\dot{B}}{B},\ \Gamma^{2}_{21}=\frac{1}{2}\frac{B^{'}}{B},\ \Gamma^{3}_{30}=\frac{1}{2}\frac{\dot{C}}{C},\ \Gamma^{3}_{31}=\frac{1}{2}\frac{C^{'}}{C}.
\end{eqnarray}
These induce the following forms of Ricci tensor:
\begin{eqnarray}
R_{00}&=&\frac{1}{4}\frac{g^2}{f^2}\frac{A^{'}B^{'}}{B}+\frac{1}{4}\frac{\dot{A}\dot{B}}{AB}+\frac{1}{4}\frac{g^2}{f^2}\frac{A^{'}C^{'}}{C}+\frac{1}{4}\frac{\dot{A}\dot{C}}{AC}+\frac{1}{2}\frac{g^2}{f^2}A^{''}\nonumber\\
&-&\frac{1}{4}\frac{g^2}{f^2}\frac{A^{'2}}{A}-\frac{1}{2}\frac{\ddot{B}}{B}+\frac{1}{4}\frac{\dot{B^2}}{B^2}-\frac{1}{2}\frac{\ddot{C}}{C}+\frac{1}{4}\frac{\dot{C^2}}{C^2}, \\
R_{10}&=&\frac{1}{4}\frac{A^{'}B^{'}}{AB}+\frac{1}{4}\frac{A^{'}\dot{C}}{AC}-\frac{1}{2}\frac{\dot{B}^{'}}{B}+\frac{1}{4}\frac{B^{'}\dot{B}}{B^2}-\frac{1}{2}\frac{\dot{C}^{'}}{C}+\frac{1}{4}\frac{C^{'}\dot{C}}{C^2},\\
R_{11}&=&-\frac{1}{2}\frac{A^{''}}{A}+\frac{1}{4}\frac{A^{'2}}{A^2}-\frac{1}{2}\frac{B^{''}}{B}+\frac{1}{4}\frac{B^{'2}}{B^2}-\frac{1}{2}\frac{C^{''}}{C}+\frac{1}{4}\frac{C^{'2}}{C^2},\\
R_{22}&=&-\frac{1}{4}\frac{A^{'}B^{'}}{A}-\frac{1}{4}\frac{f^2}{g^2}\frac{\dot{A}\dot{B}}{A^2}+\frac{1}{4}\frac{f^2}{g^2}\frac{\dot{B}\dot{C}}{AC}+\frac{1}{2}\frac{f^2}{g^2}\frac{\ddot{B}}{A}-\frac{1}{4}\frac{f^2}{g^2}\frac{\dot{B}^2}{AB}-\frac{1}{4}\frac{B^{'}C^{'}}{C}+\frac{1}{4}\frac{B^{'2}}{B}-\frac{1}{2}B^{''},\\
R_{33}&=&-\frac{A^{'} C^{'}}{4 A}-\frac{f^2 \dot{A}\dot{C}}{4 A^2 g^2}+\frac{f^2 \dot{B}\dot{C}}{4 A B g^2}+\frac{f^2 \ddot{C}}{2 A g^2}-\frac{f^2 \dot{C}^2}{4 A C g^2}-\frac{B^{'} C^{'}}{4 B}+\frac{C^{'2}}{4 C}-\frac{1}{2} C^{''}.
\end{eqnarray}
The   Ricci scalar is this case is given by
\begin{eqnarray}
R&=&\frac{g^2}{2}\frac{A^{'}B^{'}}{AB}+\frac{g^2}{2}\frac{A^{'}C^{'}}{AC}+g^2\frac{A^{''}}{A}+\frac{f^2}{2}\frac{\dot{A}\dot{B}}{A^2B}+\frac{f^2}{2}\frac{\dot{A}\dot{C}}{A^2C}-\frac{g^2}{2}\frac{A^{'2}}{A^2}-\frac{f^2}{2}\frac{\dot{B}\dot{C}}{ABC}-f^2\frac{\ddot{B}}{AB}\\ && \nonumber +\frac{f^2}{2}\frac{\dot{B}^2}{AB^2}-f^2\frac{\ddot{C}}{AC}+\frac{f^2}{2}\frac{\dot{C}^2}{AC^2}+\frac{g^2}{2}\frac{B^{'}C^{'}}{BC}+g^2\frac{B^{''}}{B}-\frac{g^2}{2}\frac{B^{'2}}{B^2}+g^2\frac{C^{''}}{C}-\frac{g^2}{2}\frac{C^{'2}}{C^2}.
\end{eqnarray}

%%%%%%%%%%%%%%%%%%%%%%%%%%%%%%%%%%%%%%%%%%%%%%%%%%%%%%%%%%%%%%%%%%%%

\end{document}